\documentclass[fleqn,usenatbib]{mnras}
\usepackage{newtxtext,newtxmath}
\usepackage[T1]{fontenc}
\usepackage{graphicx}
\usepackage{amsmath}
\DeclareRobustCommand{\VAN}[3]{#2}
\let\VANthebibliography\thebibliography
\def\thebibliography{\DeclareRobustCommand{\VAN}[3]{##3}\VANthebibliography}

\newcommand{\mfemu}{\texttt{MFEmulator}} 
\newcommand{\Data}{\mathcal{D}}
\newcommand{\gp}{\textsc{gp}}
\newcommand{\normal}{\mathcal{N}}
\newcommand{\GP}{\mathcal{GP}}
\newcommand{\thetavec}{\boldsymbol{\theta}}  

\newcommand{\yvec}{\boldsymbol{y}}
\newcommand{\outputFunction}{f}
\newcommand{\outputVector}{\yvec}            

\newcommand{\Kvec}{\boldsymbol{\mathrm{K}}}  
\newcommand{\kvec}{\boldsymbol{k}}           
\newcommand{\Lya}{Lyman-$\alpha$}

\title[Multi-Fidelity Lyman-$\alpha$ Forest Emulator]{A Multi-Fidelity Emulator for the Lyman-$\alpha$ Forest Flux Power Spectrum}
\author[Fernandez et al]
{
M.A. Fernandez,$^{1}$\thanks{E-mail:mfern027@ucr.edu}
Ming-Feng Ho,$^{1}$\thanks{E-mail:mho026@ucr.edu}
and Simeon Bird$^{1}$\thanks{E-mail:simeon.bird@ucr.edu}
\\
$^{1}$Department of Physics and Astronomy, University of California Riverside, 900 University Ave, Riverside, CA 92521
}
\date{Accepted XXX. Received YYY; in original form ZZZ}
\pubyear{2022}

\begin{document}
\label{firstpage}
\pagerange{\pageref{firstpage}--\pageref{lastpage}}
\maketitle

\begin{abstract}
In this work we extend our recently developed multi-fidelity emulation technique to the simulated Lyman-$\alpha$ forest flux power spectrum.
Multi-fidelity emulation allows interpolation of simulation outputs between cosmological parameters using many cheap low-fidelity simulations and a few expensive high-fidelity simulations.
Using a test suite of small box ($30$ Mpc/h) simulations, we show that multi-fidelity emulation is able to reproduce the Lyman-$\alpha$ forest flux power spectrum well, achieving an average accuracy when compared to a test suite of $0.8\%$.
We further show that it has a substantially increased accuracy over single-fidelity emulators, constructed using either the high or low-fidelity simulations only.
In particular, it allows the extension of an existing simulation suite to smaller scales and higher redshifts.
\end{abstract}


\section{Introduction}\label{sec:intro}

The modern and future testbed for cosmology lies in small scales and non-linear structures.
Cosmological analyses exploit observations of these scales to explore questions such as the nature of dark matter, the total neutrino mass, and the thermal history of the intergalactic medium (IGM).
One of the most powerful probes of small scale structure is the Lyman-$\alpha$ forest, a series of absorption features in the spectrum of quasars \citep{1965ApJ...142.1633G, 1998MNRAS.301..478T, 2000ApJ...543....1M, 2001ApJ...552...15H, 2002MNRAS.329..848V, 2006AJ....132..117F, 2006MNRAS.365..231V, 2006ApJS..163...80M}.
Numerical simulations are required to analyze these observations as they probe the non-linear regime.
As these simulations are expensive, cosmologists build emulators \citep{Heitmann:2006,Habib:2007,Heitmann:2009}, which interpolate a summary statistic (in this case the 1D \Lya~forest flux power spectrum) between simulation outputs at different cosmological parameters.

A recent development in cosmology is the application of multi-fidelity emulators, which allow simulations with different particle loads, and thus costs, to be combined together \citep{2022MNRAS.509.2551H}.
Here we adapt the multi-fidelity emulation technique to the \Lya~forest 1D flux power spectrum.
Multi-fidelity emulation is especially useful in this context because the \Lya~forest probes a range of redshifts, and is sensitive to smaller scales, which require higher resolution simulations, at higher redshifts \citep{2009MNRAS.398L..26B}.
The models developed here will allow a single emulator to target the wide range of scales probed by the Lyman-$\alpha$ forest, which would otherwise require a computationally infeasible number of very large simulations \citep{2014JCAP...07..005B}.

The Lyman-$\alpha$ forest is the result of overlapping neutral hydrogen absorption profiles in the spectra from distant luminous quasars, processed through the expansion of the universe \citep{1965ApJ...142.1633G}.
As light travels from the quasar, it passes through neutral hydrogen gas of varying densities.
In the rest frame of those neutral hydrogen islands, light that has been redshifted close to the Lyman-$\alpha$ transition at $1215.67${\AA} will be absorbed and the rest transmitted.
This is repeated as the light continues to intersect more neutral hydrogen islands on its path towards us, the observers.
The result is a quasar transmission spectra containing an overlapping field of absorption features that provides a proxy to the dark matter density along that sightline \citep{1998ApJ...495...44C}.

The densities probed by the Lyman-$\alpha$ forest, from redshift $2-5$, are $\sim 1-100 \ \times$ the cosmological mean density.
At these densities stellar winds and star formation effects are negligible, although black hole feedback is important \citep{2013MNRAS.429.1734V, 2020MNRAS.495.1825C}.
The densities along with the range of scales accessed has made the Lyman-$\alpha$ forest popular in cosmological studies, including: constraining the thermal history of the IGM and thus reionization \citep{2008MNRAS.386.1131B,2014MNRAS.438.2499B, 2016MNRAS.463.2335N, 2019ApJ...872..101B, 2019MNRAS.490.3177W,2021MNRAS.506.4389G, 2021arXiv211100019V}, constraining cosmological parameters including the neutrino mass \citep{2004MNRAS.354..684V, 2005ApJ...635..761M, 2006MNRAS.370L..51V, 2005PhRvD..71j3515S, 2006JCAP...10..014S, 2020JCAP...04..038P, 2021JCAP...03..049G}, and testing alternatives to cold dark matter \citep{2005PhRvD..71f3534V,  2013PhRvD..88d3502V, 2017PhRvD..96b3522I, 2020JCAP...04..038P, 2021MNRAS.502.2356G, 2021PhRvL.126g1302R}.

The Lyman-$\alpha$ forest 1D flux power spectrum is the most commonly used summary statistic for \Lya~forest spectra.
It probes small scale structure by measuring the two-point Fourier-space correlation between neutral hydrogen absorption within a sightline \citep{1998ApJ...495...44C}.

Current observational measurements of the Lyman-$\alpha$ forest flux power spectrum come from either a lower resolution, larger sample survey (SDSS, \cite{2019JCAP...07..017C}), or various higher resolution, smaller sample surveys \citep{2017MNRAS.466.4332I, 2022MNRAS.509.2842K, 2019MNRAS.489.2536D}.
In \cite{2019JCAP...07..017C}, the flux power spectrum constructed from BOSS and eBOSS spectra accesses redshifts from $z=2.2-4.6$ ($6\%$ \& $18\%$ average uncertainty, respectively) and scales from $k\approx0.001-0.02$ km$^{-1}$ s ($6\%$ \& $14\%$ average uncertainty, respectively).
The small sample, higher resolution surveys generally access a similar redshift range, but shift both the largest and smallest scales to higher $k$.
For example, in \cite{2022MNRAS.509.2842K} (their conservative results), using spectra from multiple surveys (XQ-100, KODIAQ, and SQUAD) they access redshifts from $z=2-4.6$ ($7\%$ \& $27\%$ average uncertainty, respectively) and scales from $k\approx0.005-0.1$ km$^{-1}$ s ($12\%$ \& $9\%$ average uncertainty, respectively).

The Dark Energy Spectroscopic Instrument (DESI) will soon report its first year results.
Ultimately it will increase the number of \Lya~quasar spectra by a factor of four over SDSS.
This corresponds to  $\sim 50$ quasars per square degree and a total of $7\times10^5$ quasars over the $14,000$ square degree survey footprint \citep{2016arXiv161100036D}.
In addition, DESI is expected to measure the 1D flux power spectrum at smaller scales ($k < 0.035$ km$^{-1}$ s) and higher redshifts ($z>4.6$) than SDSS, achieving order of a few percent accuracy \citep{2022arXiv220307491V}.

Extracting cosmological information from these observations will require simulations which follow the distribution of gas at relevant densities and on relevant scales.
For the \Lya~forest, box sizes of at least $100$ Mpc h$^{-1}$ and mean particle spacing of $100/3072 \approx 0.03$ Mpc h$^{-1}$ are necessary \citep{2014JCAP...07..005B}.
Earlier work has focused on methods which can reduce the cost of such simulations.
\cite{2014JCAP...07..005B} used a splicing technique to produce high resolution, large volume outputs from three sets of less computationally intensive simulations: low resolution, large volume; high resolution, small volume; and low resolution, small volume.
\cite{2015MNRAS.446.3697L} explored the use of Richardson extrapolation to enhance output resolution, in addition to testing the splicing technique.

Parameter inference tasks, such as a direct Markov Chain Monte Carlo analyses, require $\sim10^5-10^6$ model evaluations, indicating the number of simulations required by a naive approach.
Even using techniques such as splicing, this is computationally infeasible.
However, using a significantly reduced number of simulations ($\sim30$), an emulator can be constructed that effectively interpolates between this smaller set of simulations.
In addition to the \Lya~forest, emulators have been used extensively in cosmology for studying: the matter power spectrum \citep{Heitmann:2009, Heitmann:2014, Lawrence:2017, Giblin:2019, Euclid:2021, Arico:2021, Giri:2021}, weak lensing \citep{Harnois:2019, Davies:2021}, the halo mass function \citep{McClintock:2019, Nishimichi:2019, Bocquet:2022}, and the 21-cm signal \citep{Kern:2017, Cohen:2020, Bevins:2021, Bye:2022}.
Still, the computational resources required to run $\sim30$ simulations with the requisite volume and resolution is highly restrictive, especially for the \Lya~forest.

Here, we use modern machine learning techniques to alleviate the computational resource cost associated with constructing an emulator.
Specifically, we are concerned with using machine learning to predict high resolution simulation outputs to a high degree of accuracy, while running a minimal number of high resolution simulations.
One machine learning method that is suited to this task is a Gaussian Process (GP) emulator.
Gaussian processes \citep{2006gpml.book.....R} are a means of interpolating between the simulation outputs, providing function prediction in a Bayesian framework.
Essentially, a distribution of functions is learned through training on simulations, and the mean (best estimate) and variance (interpolation error) of the output can be returned for arbitrary simulation inputs.
While other interpolation methods are possible, Gaussian processes have many benefits: the inherent quantification of prediction uncertainty, the option to incorporate prior knowledge, and the ability to interpolate within high-dimensional parameter space.

Previous uses of GP emulators for the Lyman-$\alpha$ forest have been shown to be effective at predicting summary statistics \citep{2019JCAP...02..050B, Rogers:2019, 2021JCAP...05..033P, 2021JCAP...04..059W, Rogers:2021a,2021PhRvL.126g1302R}.
\cite{2019JCAP...02..050B} self-consistently showed that the predicted flux power from their GP emulator (trained with $21$ simulations) agreed to within $1-2\%$ of the corresponding simulation flux power spectrum.
These GP emulators still require a substantial computational cost, as the full simulation suite must be run with sufficient volumes and resolutions for the Lyman-$\alpha$ forest.

Recently, \cite{2022MNRAS.509.2551H} implemented a multi-fidelity GP emulator \citep{10.1093/biomet/87.1.1} for the matter power spectrum.
Here, we combine and expand on the methods outlined in \cite{2019JCAP...02..050B} and \cite{2022MNRAS.509.2551H}, to produce a multi-fidelity GP emulator for the Lyman-$\alpha$ forest flux power spectrum.
In our multi-fidelity model, the training simulations are split into two fidelities; a large sample of low resolution simulations (low fidelity, LF), and a small subset of these simulations run at higher resolution (high fidelity, HF).
Note that we use fidelity and resolution interchangeably throughout this work.
Using these two training sets, the multi-fidelity emulator is trained to predict the 1D flux power spectrum that would be output by a \textit{high} resolution simulation for arbitrary cosmological and astrophysical parameters.

A multi-fidelity emulator allows us to replace some of the HF simulations that would be needed in a single-fidelity emulator with LF simulations.
This can dramatically reduce the computational cost of constructing an emulator, while retaining predictive power across parameter space.
Using this method, emulators can be constructed that make use of the full range of scales and redshifts probed by Lyman-$\alpha$ forest observations.
This enables analyses which can jointly constrain thermal, astrophysical, and cosmological parameters.

In our high resolution simulations, the Lyman-$\alpha$ forest flux power spectrum is converged to $\approx5\%$.
While the scales we probe in this work are resolved, the box size we use is smaller than required to analyze the full range of scales available in Lyman-$\alpha$ forest data, i.e. we cannot compute a likelihood function using all the real data without larger boxes.
We therefore defer a full cosmological likelihood analysis to future work.
Our goal is to quantitatively test the accuracy of the emulator output, as compared with the output from a set of testing simulations run at the same resolution as the HF training set, and demonstrate the validity and utility of the multi-fidelity technique.
Specifically, we quantify the accuracy of the single- and multi-fidelity emulators with respect to true values from the testing simulations, thus determining how effective the multi-fidelity model is at producing high resolution outputs at minimal computational cost.
\section{Simulations}\label{sec:sims}

\begin{figure*}
    \centering
	\includegraphics[width=2\columnwidth]{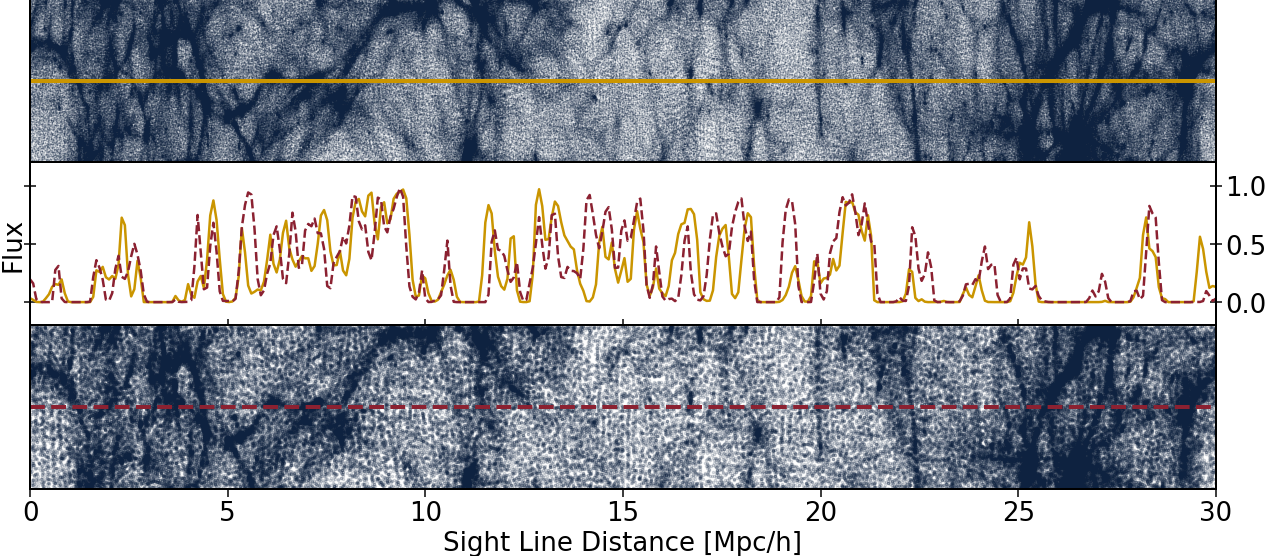}
    \caption{Example Lyman-$\alpha$ forest spectra and corresponding gas density from simulations at redshift $4$.
    The top and bottom panel show the simulated gas surrounding the skewer which produced the spectra shown in the middle panel.
    Examples are shown for simulations run at high (top panel, yellow line) and low resolution (bottom panel, red line).}
    \label{fig:spec_sim}
\end{figure*}

Simulations were performed using MP-Gadget\footnotemark, an N-body and smoothed particle hydrodynamics (SPH) code built on the solid base of Gadget-3 \citep[last described in][]{2005MNRAS.364.1105S}.\footnotetext{\url{https://github.com/MP-Gadget/MP-Gadget}}
MP-Gadget has been substantially modified to include shared-memory parallelism using OpenMP, together with many other algorithmic improvements and new subgrid models, as described in \cite{2022MNRAS.512.3703B,2022MNRAS.513..670N,2020JCAP...06..002B}.
The initial power spectrum and transfer functions are generated with the Boltzmann code CLASS \citep{2011arXiv1104.2932L}.
Species-specific initial conditions are generated for baryons and dark matter \citep{2020JCAP...06..002B, 2021MNRAS.503.1668F}.
We include radiation in the cosmological background model and assume massless neutrinos.

The physics models largely follow those used for the \texttt{ASTRID} simulation, which are described in \cite{2022MNRAS.512.3703B}.
Primary sources for these models, as well as changes from \cite{2022MNRAS.512.3703B} are described below.

We use a cubic kernel for our density estimator (rather than a quintic kernel).
For these simulations, the cubic kernel, in addition to running faster, produced a neutral hydrogen column density distribution that was more consistent with observations for column densities between $10^{20}$ and $10^{22}$ cm$^{-2}$ \citep{2021MNRAS.507..704H}.
We continue to use the pressure-entropy formulation of SPH.
The smaller SPH kernel increases the noise within galaxies, but has minimal effect on the Lyman-$\alpha$ forest \citep{2013MNRAS.429.3341B}.

Star formation follows the model of \cite{2003MNRAS.339..289S}, with our specific implementation as described in \cite{2016MNRAS.455.2778F}.
We lower the number of stars produced per gas particle from $4$ (used in \texttt{ASTRID}) to $1$ (as in Illutris-TNG), which speeds up the simulation without having an effect on the Lyman-$\alpha$ forest.

\begin{figure}
    \centering
	\includegraphics[width=\columnwidth]{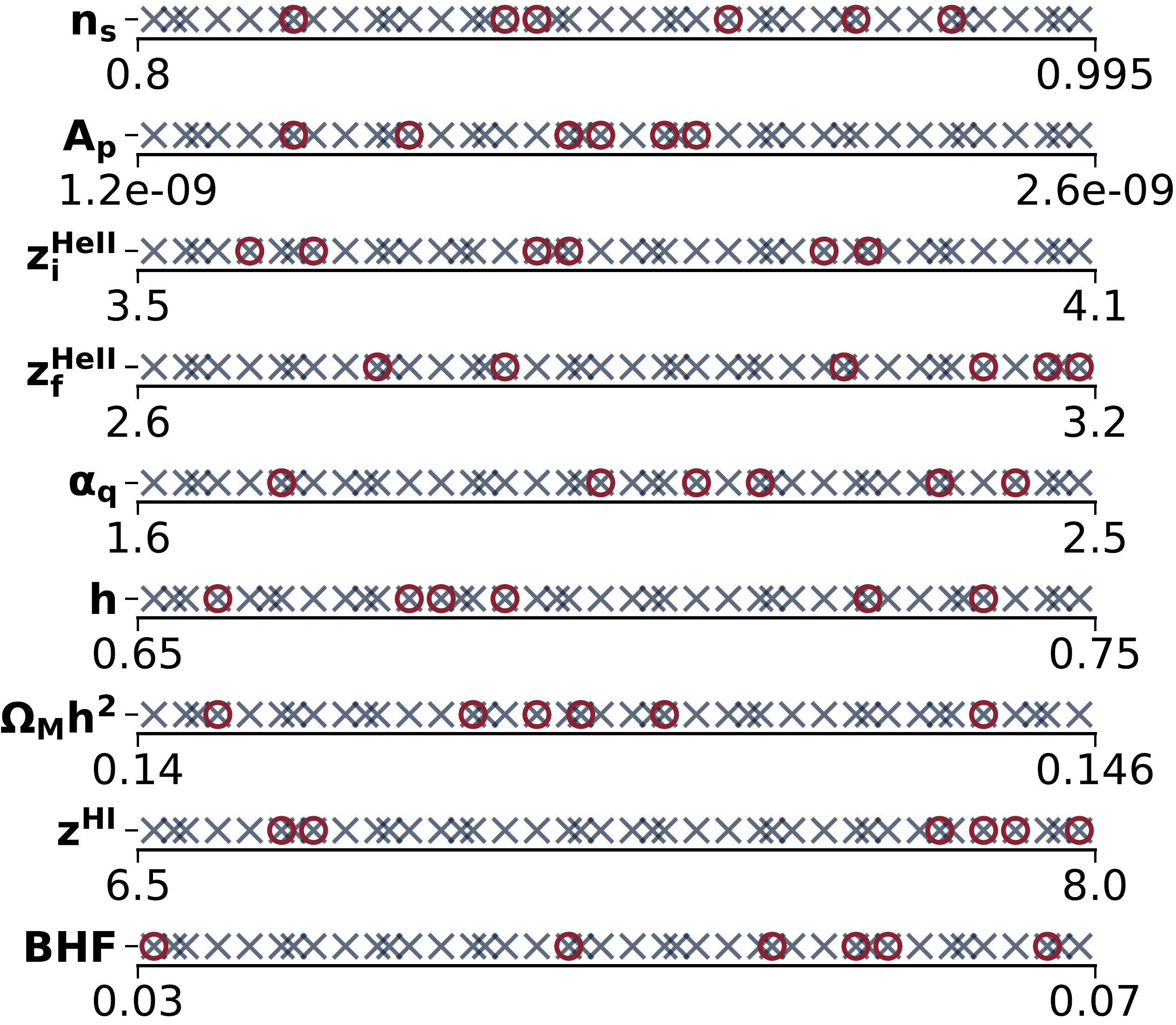}
    \caption{Simulation parameter limits and samples.
    Parameters for the low resolution simulations (crosses) were determined by filling a Latin hypercube.
    Initially, $30$ low resolution samples were generated, then an additional $10$ were added while maintaining the Latin hypercube method, hence the non-uniform spacing for the low resolution samples.
    The optimal subset of low resolution simulations is determined (see Section~\ref{subsubsec:sampling_strategy}) and this subset is run at higher resolution (shown as red circles).
    }
    \label{fig:samples}
\end{figure}

Black holes follow the model of \cite{2022MNRAS.513..670N}.
We found that for the resolutions used here, the dynamic friction from gas led to a few black holes escaping from their dark matter halo, so we only use dynamic friction from dark matter and stars.
The black hole feedback factor, which controls the fraction of luminosity that is converted to thermal energy, is an emulator parameter (BHF), with the associated parameter limits in Figure~\ref{fig:samples}.
The black hole feedback radius is fixed to $3$ kpc h$^{-1}$, selected to be the average black hole feedback radius at the highest tested resolution when using a nearest neighbour distance.
To accommodate a lower mass resolution than \texttt{ASTRID}, the minimum stellar mass needed in a halo to seed a black hole was increased to $2\times10^{8} M_{\odot}$, and all black hole seeds start with a mass of $5\times10^{4} M_{\odot}$.

Stellar winds are modeled following \cite{2010MNRAS.406..208O}.
The decoupling distance for the winds is increased from $20$ kpc h$^{-1}$ to $1$ Mpc h$^{-1}$, which allows the winds to recouple due to density changes rather than travel distance.
The density threshold for wind recoupling is set to $10\%$ of the star formation density threshold (which is $57.7$ times the critical density).
The minimum wind velocity is set to $100$ km/s.
Finally, metal return (gas enrichment) is disabled as it is not important for the Lyman-$\alpha$ forest and can be computationally expensive.

Gas is assumed to be in ionization equilibrium with a uniform ultraviolet background using the model of \cite{2020MNRAS.493.1614F}.
We boost the temperature of the gas to $15000$ K the timestep after the gas is reionized, to model impulsive heating during hydrogen reionization from ionization fronts \citep{2019ApJ...874..154D}.

We implement He~{\sc ii} reionization using the model of \cite{2020MNRAS.496.4372U}.
The input parameters for this model are: quasar mean bubble size and variance, redshifts for the start and completion of He~{\sc ii} reionization z$^{\text{He~{\sc ii}}}_i$, z$^{\text{He~{\sc ii}}}_f$, and the quasar spectral index $\alpha_q$ (which effectively scales the peak temperature during He~{\sc ii} reionization).
The quasar bubble size is reduced from the default of $\sim30$ Mpc, motivated by radiative transfer simulations \cite{2009ApJ...694..842M}, to $5$ Mpc, due to our small box size.

Simulations are initialised at $z=99$ and finish at $z=2$, and use periodic boundaries.
Box volume, particle number, and gas particle mass resolution are reported in Table~\ref{table:simulations}.
The range given for the gas resolution is due to the varying value of $h$ in our simulation suite.
The gas particle mass resolution for our HF simulations does not meet the resolution that \citet{2009MNRAS.398L..26B} recommend to resolve the forest at all redshifts of interest.
However, the Lyman-$\alpha$ forest flux power spectrum from our HF simulations is converged to within $\approx5\%$ of a simulation that does meet the required resolution of \cite{2009MNRAS.398L..26B}.
We are interested in the performance of the multi-fidelity GP emulator in learning the mapping from low to high resolution, thus this slight lack of numerical convergence does not affect our results.
Examples of the gas density (at $z=3.6$) for the two resolutions are shown in the top and bottom panels of Figure~\ref{fig:spec_sim}.

\begin{table}
	\centering
	\caption{Table of simulation sets}
	\label{table:simulations}
	\begin{tabular}{lcccc}
		\hline
		Simulation & Box Volume & N$_{\text{part}}$ & M$_{\text{gas}}$ (M$_{\odot}$ h$^{-1}$)\\
		\hline
		LF & $(30$ Mpc h$^{-1})^3$ & $2\times256^3$ & $[1.78, 2.37]\times10^7$\\
		HF & $(30$ Mpc h$^{-1})^3$ & $2\times512^3$ & $[2.22, 2.96]\times10^6$\\
		Test & $(30$ Mpc h$^{-1})^3$ & $2\times512^3$ & $[2.22, 2.96]\times10^6$\\
		\hline
	\end{tabular}
\end{table}

\begin{figure}
    \centering
	\includegraphics[width=\columnwidth]{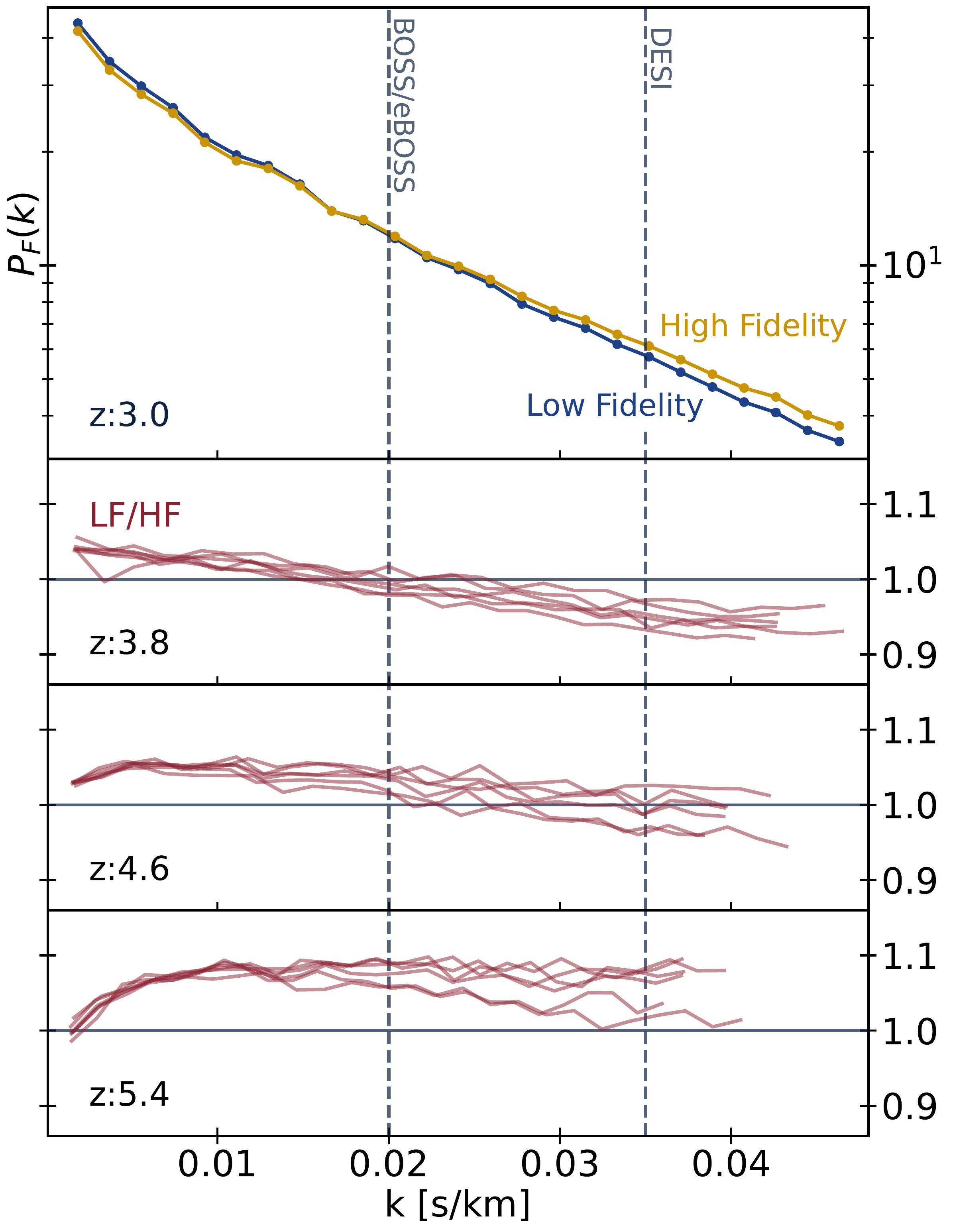}
    \caption{
    Lyman-$\alpha$ forest flux power spectrum from LF and HF simulations.
    The top panel shows the flux power spectrum at redshift $3$ from an LF simulation (blue), and its HF counterpart (yellow).
    The lower panels show the ratio of the flux power for these two resolutions, for all LF-HF simulation pairs, at $z=3.8, 4.6, \& \ 5.4$.
    Dashed lines shown the highest $k$ probed by BOSS/eBOSS \citep{2019JCAP...07..017C}, and the estimated reach for DESI \citep{2022arXiv220307491V}.}
    \label{fig:low_high_comp}
\end{figure}

Lyman-$\alpha$ forest absorption spectra are generated using Fake Spectra Flux Extractor \citep{2017ascl.soft10012B}\footnotemark, described in \citet{2015MNRAS.447.1834B}.
\footnotetext{\url{https://github.com/sbird/fake_spectra}}
We generate $32,000$ (seeded) randomly placed skewers for each snapshot, from $z=5.4$ to $z=2.0$ in increments of $\Delta z=0.2$.
The pixel resolution is set to $10$ km s$^{-1}$.
An optical depth threshold of $\tau < 10^6$ is set to eliminate damped Lyman-$\alpha$ systems.
Example spectra from one LF and one HF simulation (at $z=4$) are shown in the middle panel of Figure~\ref{fig:spec_sim}.
Note that the LF simulation is not simply a smoothed version of the HF simulation, as the fine velocity structure of the gas moves the location of the absorption peaks.

These sets of neutral hydrogen absorption spectra are used to construct the Lyman-$\alpha$ forest flux power spectrum for each simulation, at each redshift.
The flux power spectrum is defined as $P_F(k) = |L^{-1}\tilde{\delta}^2_F(k)|$, where $\tilde{\delta}^2_F(k)$ is the Fourier transform of the flux excess, $\delta_F(k) = F(k)/\langle F(k) \rangle - 1$, and $L$ is the length of the sightline.
The reported flux power spectrum is averaged over all $32,000$ spectra.

Figure~\ref{fig:low_high_comp} shows flux power spectra from a single LF simulation and its HF counterpart, and the ratio of these at several redshifts.
While the exact difference between the LF and HF flux power spectra depends on simulation input parameters and redshift, in general the LF differs most from the HF at small scales.
The enhanced power on large scales for the LF flux power spectra is consistent with \citet{2014JCAP...07..005B}, and is likely due to differences in heating and cooling during H~{\sc i} and He~{\sc ii} reionization.

Figure~\ref{fig:samples} lists the input parameters that are varied across our suite of simulations, as well as their limits.
Two parameters control the primordial power spectrum: $n_s$ is the scalar spectral index (slope) and $A_p$ is the amplitude (see \cite{2019JCAP...02..050B} for more details).
Three of the parameters relate to the He~{\sc ii} reionization model: z$^{\text{He~{\sc ii}}}_i$ and z$^{\text{He~{\sc ii}}}_f$ are the redshifts for the start and end of He~{\sc ii} reionization, and $\alpha_q$ is the quasar spectral index.
We vary the Hubble constant through $h$, and the total matter density through $\Omega_M h^2$.
One parameter is varied for H~{\sc i} reionization: z$^{\text{H~{\sc i}}}$ is the midpoint redshift of H~{\sc i} reionization.
Finally, we vary one parameter for the black hole model: $BHF$ is the black hole feedback factor, which controls the fraction of luminosity that is converted to thermal energy.
Note that our simulations do not contain a kinetic feedback model.
However, at $z > 2$ it is expected that the thermal mode dominates.
Also shown in Figure~\ref{fig:samples} are the LF and HF training samples.
Note that the HF samples are a subset of the LF samples.
The selection of the HF samples is described in Section~\ref{subsubsec:sampling_strategy}.

\section{Emulators}\label{sec:emulators}

In Section~\ref{subsec:gp_emu}, we will briefly review emulation using a Gaussian process.
In Section~\ref{subsec:mfemu_emu}, we will review how the Gaussian process emulator can be extended to model simulations with different qualities using a multi-fidelity emulator, \mfemu.
The multi-fidelity emulation technique of \cite{10.1093/biomet/87.1.1} will be reviewed in Section~\ref{subsubsec:mfemu_emu_ko}.
Section~\ref{subsubsection:differences} will discuss the differences in multi-fidelity emulator design between this paper and \cite{2022MNRAS.509.2551H}.
Finally, Section~\ref{subsubsec:sampling_strategy} will outline how we select our training simulations in the parameter space.

\subsection{Gaussian process emulator}\label{subsec:gp_emu}

Gaussian process (\gp) regression models \citep{2006gpml.book.....R} have been widely used to build cosmological emulators \citep{Heitmann:2006, Habib:2007, Heitmann:2009}.
A {\gp} provides closed-form expressions for predictions.
In addition, a {\gp} naturally comes with uncertainty quantification, which is useful when building an inference framework.
In the context of emulation, a {\gp} can be seen as a Bayesian prior for the simulation response. It is a prior because the emulator model is chosen to ensure smoothness and monotonicity features of the simulation response \textit{before} data are collected \citep{Santner:2003}.

Let $\thetavec \in \Theta \subseteq \mathbb{R}^{d}$ be the input cosmologies for the simulator, where $d$ is the dimension of the parameters ($d = 9$ for our emulator).
$\outputFunction(\thetavec)$ is the corresponding output summary statistic.
In this work, the summary statistic, $\outputFunction(\thetavec)$, is the Lyman-$\alpha$ forest flux power spectrum.
A {\gp} regression model can be viewed as a prior on the response surface of our simulated Lyman-$\alpha$ forest flux power spectrum:
\begin{equation}
    \outputFunction(\thetavec) \sim \GP(\mu(\thetavec), k(\thetavec, \thetavec')),
\end{equation}
where $\mu(\thetavec) = \mathbb{E}[\outputFunction(\thetavec)]$ is the mean function, and $k(\thetavec, \thetavec') = \mathrm{Cov}[\outputFunction(\thetavec), \outputFunction(\thetavec')]$ is the covariance kernel function.
In this work, we assume a zero mean function, and we used the same covariance function as \cite{2019JCAP...02..050B}, which will be defined later in this section.

Suppose we run the simulations at $n$ carefully chosen input cosmologies, $\mathcal{D} = \{\thetavec_1, \cdots, \thetavec_n\}$, and we generate the corresponding Lyman-$\alpha$ forest flux power spectrum for each simulation, $\outputVector = \{ \outputFunction(\thetavec_1), \cdots, \outputFunction(\thetavec_n) \}$.
Conditioning on this training data, we can get the predictive distribution of $\outputFunction$ at a new input cosmology $\thetavec$ through the closed-form expression:
\begin{equation}
    \outputFunction(\thetavec) \mid \outputVector, \Data \sim \normal(\mu_n(\thetavec), \sigma_n^2(\thetavec)),
\end{equation}
where the mean and variance are:
\begin{equation}
    \begin{split}
        \mu_n(\thetavec) &= \kvec(\thetavec, \Data)^\intercal \Kvec(\Data)^{-1} \outputVector;\\
        \sigma_n^2(\thetavec) &= k(\thetavec, \thetavec) - \kvec(\thetavec, \Data)^\intercal \Kvec(\Data)^{-1} \kvec(\thetavec, \Data).
    \end{split}
\end{equation}

The vector $\kvec(\thetavec, \Data) = [ k(\thetavec, \thetavec_1), \cdots, k(\thetavec, \thetavec_n) ]$ represents the covariance between the new input cosmology, $\thetavec$, and the training data.
The matrix $\Kvec(\Data)$ is the covariance for the training data.

For the covariance kernel function, we choose the same kernel as in \cite{2019JCAP...02..050B}, which is a combination of a linear kernel and a radial basis kernel (RBF):
\begin{equation}
    \begin{split}
        k(\thetavec,\thetavec'; \sigma_0, \boldsymbol{l}, \boldsymbol{\sigma}) &= k_\mathrm{RBF}(\thetavec, \thetavec'; \sigma_0, \boldsymbol{l}) + k_\mathrm{LIN}(\thetavec, \thetavec'; \boldsymbol{\sigma})\\
        &= \sigma_0^2 \exp{\left( \sum_{i=1}^{d} -\frac{(\thetavec_i - {\thetavec_i}')^2}{2 l_i^2} \right)} +  \sum_{i=1}^{d} \sigma_i^2 \thetavec_i {\thetavec_i}',
    \end{split}
\end{equation}
where $\sigma_0^2$ and $\boldsymbol{\sigma}^2$ are the variance hyperparameters for the RBF kernel and the linear kernel, respectively.
$\boldsymbol{l}$ is the lengthscale parameter that controls the smoothness of the Gaussian process function.
We applied Automatic Relevance Determination (ARD) for both linear and RBF kernels.
That is, we assign one lengthscale $l_i$ (variance $\sigma_i$) hyperparameter for each input dimension $i$ for the RBF and linear kernels.
This allows the GP to dynamically learn the scale over which each input dimension varies, which corresponds to the degree of sensitivity of the flux power spectrum to the input parameter.

Although we do not explicitly write in the notation, $\outputFunction(\thetavec)$ is a single-valued output.
Since our target summary statistic is a vector, we model each $k$-bin of the flux power spectrum with a separate {\gp}.
The primary reason for this choice is that the correlation between the low-fidelity and high-fidelity flux power spectrum changes depending on the scale considered.
The multi-fidelity method can only capture this scale dependence if we model each scale separately.

\subsection{Multi-Fidelity Emulation}\label{subsec:mfemu_emu}

We first introduce the Kennedy-O'Hagan model (KO model) \citep{10.1093/biomet/87.1.1} in Section~\ref{subsubsec:mfemu_emu_ko}.
Section~\ref{subsubsection:differences} describes the changes we have made to adapt the model from \cite{2022MNRAS.509.2551H} to the \Lya~forest.
Finally, the strategy we employ for choosing parameters at which to generate high-fidelity training simulations is described in Section~\ref{subsubsec:sampling_strategy}.

\subsubsection{Kennedy O'Hagan Method}
\label{subsubsec:mfemu_emu_ko}

The KO model \citep{10.1093/biomet/87.1.1} was first introduced to model a sequence of computer codes with increasing fidelity.
For simplicity, we assume there are only two fidelities:
low-fidelity (LF) simulations with low resolution,
and high-fidelity (HF) simulations with high resolution.

We define $\{\outputVector_\mathrm{LF}, \outputVector_\mathrm{HF}\}$ as the {\Lya} forest flux power spectra in the training set. $\outputVector_\mathrm{LF} = \{\outputFunction_\mathrm{LF}(\thetavec^\mathrm{LF}_i) \}_{i=1}^{n_\mathrm{LF}}$  and $\outputVector_\mathrm{HF} = \{\outputFunction_\mathrm{HF}(\thetavec^\mathrm{HF}_i) \}_{i=1}^{n_\mathrm{HF}}$, where $n_\mathrm{LF}$ and $n_\mathrm{HF}$ are the number of simulations in the low- and high-fidelity training sets.
We use the KO method to model \Lya~forest flux power spectra from different fidelities:
\begin{equation}
    \outputFunction_\mathrm{HF}(\thetavec) = \rho \cdot \outputFunction_\mathrm{LF}(\thetavec)
    + \delta(\thetavec),
    \label{eq:ko_model}
\end{equation}
where $\rho$ is a trainable parameter describing a multiplicative correction between the low- and high-fidelity {\Lya} forest flux power spectra.
$\delta(\thetavec)$ is a GP independent of $\outputFunction_\mathrm{LF}(\thetavec)$, describing an additive correction between fidelities.
In other words, Equation~\ref{eq:ko_model} assumes the high-fidelity {\Lya} forest flux power can be decomposed as the low-fidelity flux power multiplied by a correction parameter, $\rho$, and an additive bias function $\delta(\thetavec)$.

\begin{figure}
    \includegraphics[width=\columnwidth]{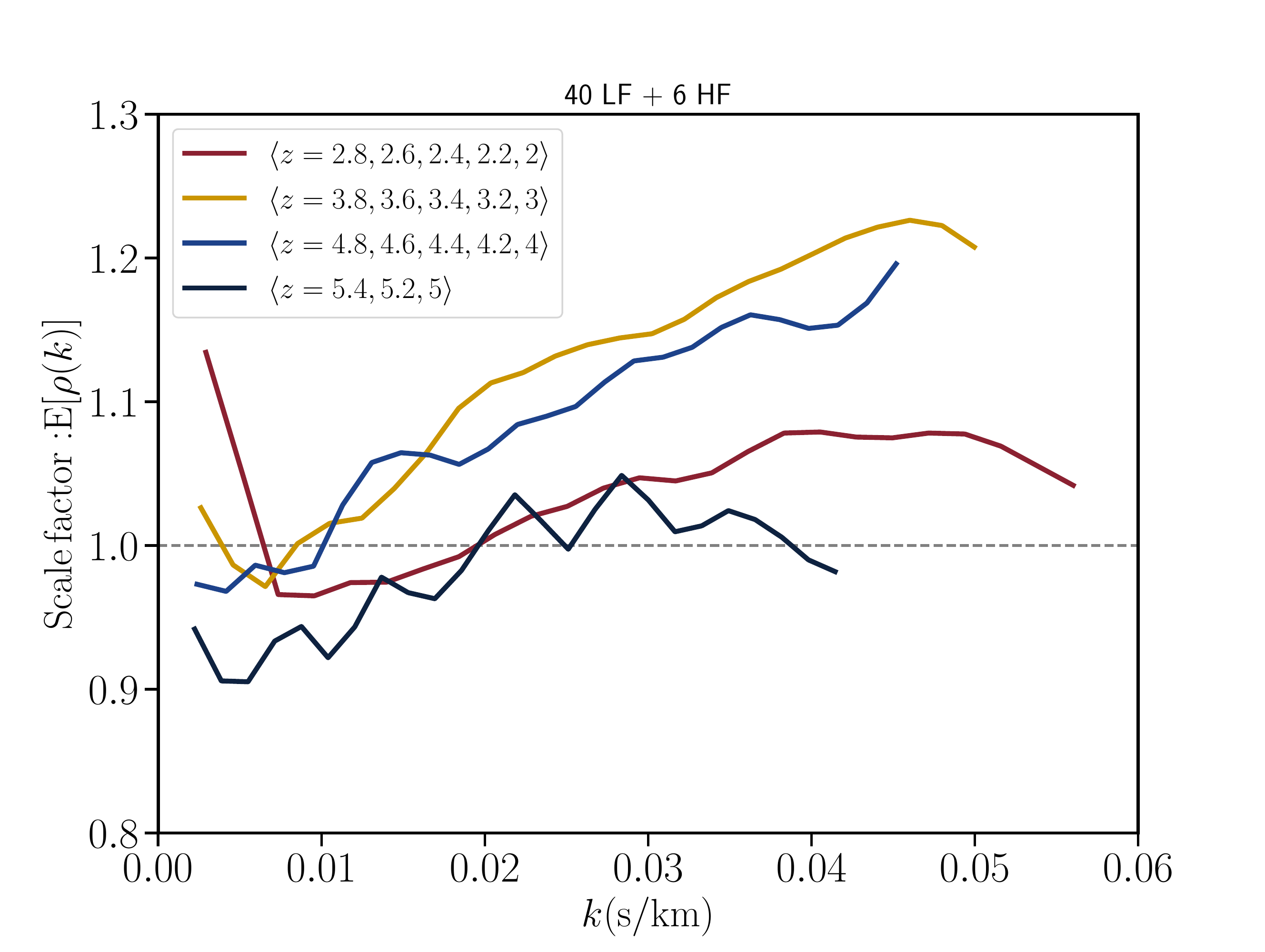}
    \caption{
    The scale parameter, $\rho$, of the KO model (Equation~\ref{eq:ko_model}) as a function of $k$.
    Different colors represent different redshifts.
    We separate the redshifts into four bins, $2 \leq z < 3$, $3 \leq z <4$, $4 \leq z < 5$, and $5 \leq z \leq 5.4$.
    Within each redshift bin, we average over fixed $k$ modes, which are linearly spaced between the maximum $k$ and minimum $k$ at the given redshift range.
    This emulator used 40 LF and 6 HF for training.
    }
    \label{fig:ko_rho}
\end{figure}

As mentioned in \cite{2022MNRAS.509.2551H}, the $\rho$ parameter has to be scale-dependent (a function of $k$) to model the well-known fact that small scales are less well resolved in smaller simulations.
Here we use the same method as \cite{2022MNRAS.509.2551H} and assume Equation~\ref{eq:ko_model} is a single-output GP model. We assign a KO model to each $k$ bin of the data.\footnote{We can easily get the same set of $k$ bins for low- and high-fidelity by using the same spectral resolution for both simulations.}
In this way, we can model $\rho$ as a function of $k$, as shown in Figure~\ref{fig:ko_rho}.

We also assign KO models for each redshift.
As shown in Figure~\ref{fig:ko_rho}, $\rho$ is a non-trivial function of both $k$ and $z$, so we cannot simply use an emulator trained on one redshift to apply on another redshift.\footnote{See \cite{2021JCAP...05..033P} for a Lyman-$\alpha$ forest emulator that uses a single GP for all redshifts, and achieves sub-percent accuracy, albeit with some ambiguity between model parameters and redshift.}
We note that it is possible to assume a smooth function to model $\rho(k, z)$.
However, validating $\rho(k, z)$ is out-of-scope for this paper.
In practice, observational data are conditioned on a specific redshift, so training emulators on separate redshifts is sufficient for cosmology inference.

Figure~\ref{fig:ko_rho} shows that $\rho$ stays close to unity at large scales for most of the redshifts.
At small scales, however, different redshifts require different values of $\rho$.
At the middle redshifts ($3 \leq z < 5$), $\rho$ has a large positive deviation from unity.
At the low redshifts ($2 \leq z < 3$), $\rho$ has a moderate deviation toward $\rho > 1$.
The only exception is at the high redshifts ($5 \leq z \leq 5.4$), which stays close to $\rho = 1$ for all scales.
This indicates that the correction to the {\Lya} forest flux power due to the resolution of the simulation varies with redshift, depending on the over-density probed by the forest.

\subsubsection{Model differences from Ho et al. (2022)}\label{subsubsection:differences}

Here we highlight the ways in which we have adapted the model from \cite{2022MNRAS.509.2551H}, which emulated the non-linear matter power spectrum at $z= 0 - 2$, to the \Lya~forest at $z=2 - 5.4$.
Since the redshift range is larger in this work, we employed a new strategy to select the optimal HF training set by averaging over the interpolation loss for all redshift bins.
We will describe the strategy in detail in Section~\ref{subsubsec:sampling_strategy}.

In \cite{2022MNRAS.509.2551H}, we outlined two multi-fidelity methods: a linear multi-fidelity emulator (the KO model, or the autoregression model (AR1)), and a non-linear multi-fidelity emulator (non-linear autoregressive GP, or NARGP, \cite{2017RSPSA.47360751P}).
However, we found that NARGP requires more HF training simulations for the {\Lya} forest flux power than AR1, perhaps due to the wide range of redshifts used.
We use the KO model for our main results, and describe the NARGP results in Appendix~\ref{sec:nonlinear}.

In this work, instead of emulating logarithm scaled powers, we adopted the mean-flux normalization strategy proposed in \cite{2019JCAP...02..050B}.
We normalize all flux power spectra in the training set by the median spectrum:
\begin{equation}
    \begin{split}
        \yvec_\mathrm{LF} &\leftarrow \frac{\yvec_\mathrm{LF}}{\mathrm{median}_i{(\yvec_\mathrm{LF})}} - 1;\\
        \yvec_\mathrm{HF} &\leftarrow \frac{\yvec_\mathrm{HF}}{\mathrm{median}_i{(\yvec_\mathrm{LF})}} - 1.        
    \end{split}
    \label{eq:normalization}
\end{equation}
The index $i$ refers to one of the spectra in the training set, $\outputVector_\mathrm{LF} = \{\outputFunction_\mathrm{LF}(\thetavec^\mathrm{LF}_i) \}_{i=1}^{n_\mathrm{LF}}$.
Equation~\ref{eq:normalization} ensures the training sample distribution is close to having a zero mean, matching the prior of the GP emulator. 
We found that in practice this normalisation makes training the emulator substantially easier. Note that we normalize the HF training set using the same LF median spectrum. As the HF training set is small, the median spectrum estimate for HF is noisy, and so using it for normalization may introduce some unwanted training bias.

\subsubsection{Sampling strategy for high-fidelity simulations}\label{subsubsec:sampling_strategy}

The KO model approach can be seen as a Bayesian way to correct an emulator from low-fidelity to high-fidelity.
Thus, if $\outputVector$ is the high-fidelity {\Lya} forest flux power spectrum used for training, and $\thetavec$ is the corresponding input parameters:
\begin{equation}
    \begin{split}
        \outputVector &= \outputFunction_\mathrm{LF}(\thetavec) + (\outputFunction_\mathrm{HF}(\thetavec) - \outputFunction_\mathrm{LF}(\thetavec))\\
              &= \outputFunction_\mathrm{LF}(\thetavec) + \mathrm{error}(\thetavec).
    \end{split}
\end{equation}
The emulation accuracy will be directly affected by how well an autoregressive construction can model $\mathrm{error}(\thetavec)$.
Usually, a large set of low-fidelity simulations are used as training data for $\outputFunction_\mathrm{LF}(\thetavec)$ because they can be obtained cheaply.
The quality of training data for $\mathrm{error}(\thetavec) = \outputFunction_\mathrm{HF}(\thetavec) - \outputFunction_\mathrm{LF}(\thetavec)$ thus relies on the choice of high-fidelity simulations.

In \cite{2022MNRAS.509.2551H}, we proposed an optimization strategy to select high-fidelity training simulations.
A low-fidelity only emulator (LFEmu)\footnote{In a similar way, we call a high-fidelity only emulator, HFEmu.} is trained on a subset of low-fidelity training simulations.
The posterior means of the trained LFEmu are used to calculate the emulation errors from the remaining LF samples in the Latin Hypercube Sampling (LHS).
By minimizing the emulation errors of LFEmu, we can grid search for the optimal set of cosmologies that best interpolates the parameter space using a small number of training simulations.
Assuming LFEmu is correlated with HFEmu, we can use the selected optimal cosmologies as inputs for the HF training set.
By ensuring the HF training set achieves a good interpolation, we mitigate emulation errors for the multi-fidelity emulator.

In practice, we employ a three-stage procedure for building a multi-fidelity emulator:
\begin{enumerate}
    \item Prepare LF simulation suite.
    \item Prepare HF simulation suite. This is done by using LFEmu to find the set of cosmologies that minimizes the interpolation loss.
    \item Build {\mfemu}. If the accuracy is not enough, go back to stage 1 or 2 to run more training simulations.
\end{enumerate}
For stage (ii), to avoid wasting computational resources running more LF simulations, we directly use the LF simulation suite in stage (i) to build and validate the LFEmu.
Thus, the cosmologies chosen for the HF set are a subset of the LF simulation LHS, which fulfills the nested training dataset design suggested in \cite{10.1093/biomet/87.1.1}.
The benefit of using a nested data structure, $\thetavec_\mathrm{HF} \subseteq \thetavec_\mathrm{LF}$, is that we can directly compute posterior means from the LF training set for cosmologies $\thetavec_\mathrm{HF}$, without any interpolation in LF.

We note that it is possible to train a {\mfemu} without using the LF simulations to optimize the HF points.
However, if the selection of HF points are suboptimal (i.e., can barely interpolate in the prior volume), then the {\mfemu} accuracy will be suboptimal.
This is because the $\mathrm{error}(\theta)$ cannot be decomposed into an autoregressive structure easily.

To find the optimal HF training set across the full redshift range, $z = 2 - 5.4$, we train a LFEmu for each redshift and get the validation loss (we used mean squared errors).
We sum up the validation loss for all redshifts and find a subset of cosmologies that minimizes the summed validation loss.

\begin{figure}
    \includegraphics[width=\columnwidth]{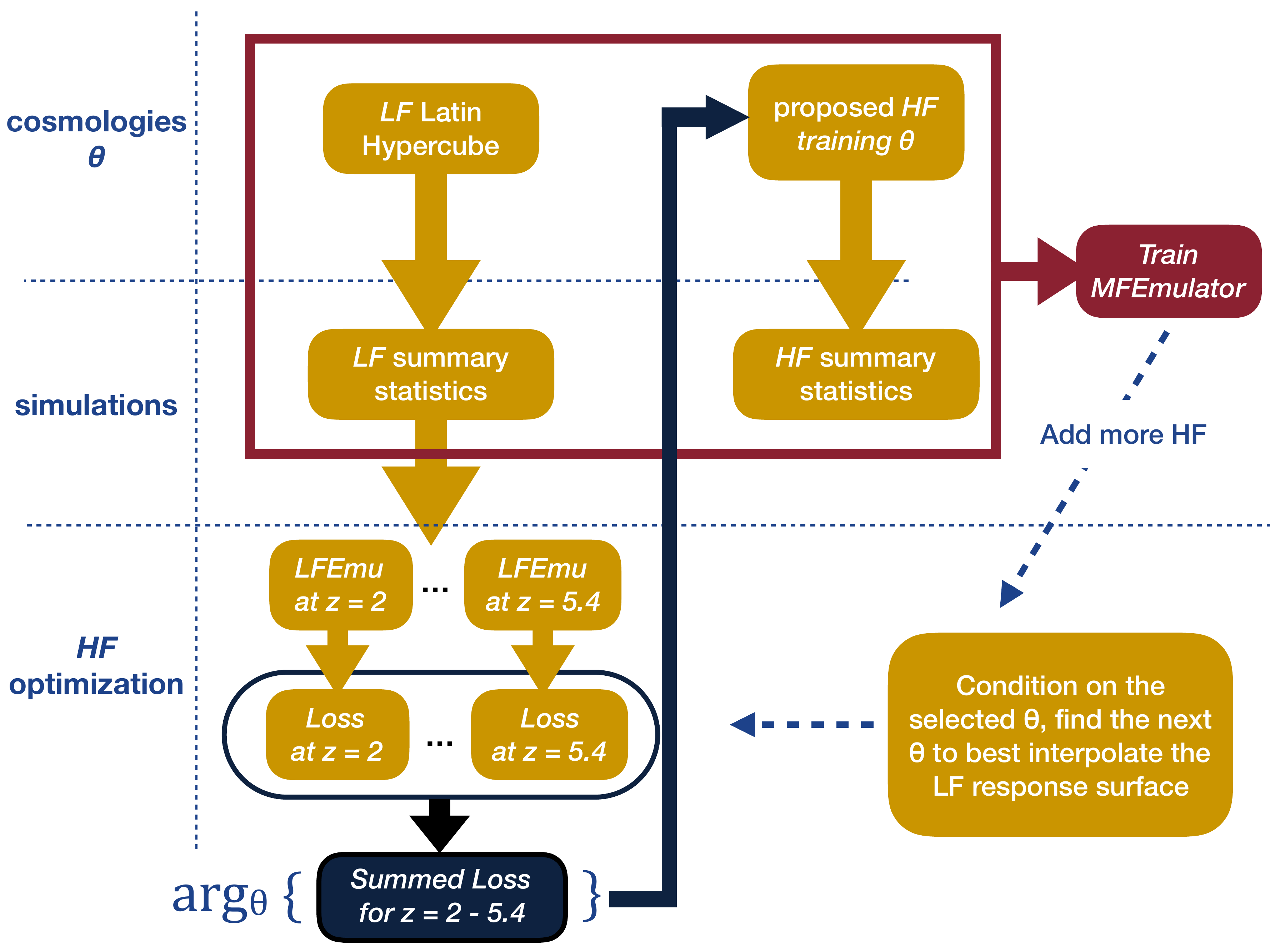}
    \caption{A flowchart for training a multi-fidelity emulator.
    We start with a low-fidelity (LF) set of simulations.
    We then select a subset of low-fidelity points to train an emulator, LFEmu, at each redshift.
    The validation loss for each LFEmu is computed using the rest of the LF simulations as a validation set.
    Finally, we sum up the validation losses at each redshift, and use the summed loss to propose a set of cosmologies $\thetavec_\mathrm{HF}$ which can best minimize the summed loss.}
    \label{fig:hf_optimize_flowchart}
\end{figure}

In Figure~\ref{fig:hf_optimize_flowchart}, we summarize the above described procedure in a flowchart.
For a formal description, we refer to \cite{2022MNRAS.509.2551H}.
The only difference between the proposed procedure in Figure~\ref{fig:hf_optimize_flowchart} and the procedure in \cite{2022MNRAS.509.2551H} is that we optimize $\thetavec_\mathrm{HF}$ across redshifts $z = 2 - 5.4$ in this work.
Thus, we have an additional step to sum the LFEmu validation loss across redshifts.

Training an LFemu on all possible low-fidelity subsets is computationally intensive.
To reduce costs, we employed the greedy optimization strategy from \cite{2022MNRAS.509.2551H}.
We first explored all possible subsets for $3$ design points within the LF LHS.
For the optimal $4$ design points, instead of exploring all possible subsets, we grew the subset one point at a time, fixing the previously chosen optimal $3$ HF points.
In the same line of thought, we grew the subset to $6$ optimal design points for HF training cosmologies.
Our final simulation suite of $40$ LF and $6$ HF samples, along with parameter limits, is shown in Figure~\ref{fig:samples}.

\section{Results \& Discussion}\label{sec:results}

Using the flux power spectra from our LF and HF simulations, we train single-fidelity (one LF only, and one HF only) and multi-fidelity emulators.
These trained emulators are used to predict the flux power spectrum output for a set of $10$ simulation input parameters.
We then compare these predictions to the corresponding testing simulations which were run at the same resolution as the HF simulations (see Table~\ref{table:simulations}).

\subsection{Emulator Accuracy}\label{sec:accuracy}

In the following, we only show results for the emulators that use the full available set of training simulations ($40$ LF and $6$ HF).
We have verified that using all available training simulations leads to the most accurate emulator.
Section~\ref{sec:runtime} shows how emulator accuracy degrades when a smaller subset of the available simulations is used.

Using the full set of available simulations, the mean prediction error for the multi-fidelity emulator is $\langle| P_F^{\text{pred}}/P_F^{\text{true}} - 1 |\rangle \approx0.8\%$ (averaging across all scales, redshifts, and testing simulation outputs).
For the LF single-fidelity emulator, the mean prediction error is $\approx4\%$.
This is not unexpected; there are real differences between the flux power spectra output by the low and high resolution simulations that are not being captured with this method.
The $4\%$ error may seem quantitatively quite good, considering the simpler methodology and reduced resource cost.
However, there is no indication that the error could be reduced further with additional simulations (see Section~\ref{sec:runtime}).

For the HF single-fidelity emulator, the mean prediction error is $\approx3\%$.
This is likely limited by the sample size of the training set ($6$ simulations), leading to increased errors when making predictions for inputs that are far away from the training samples.
It is important to note that the HF samples are selected to optimize the multi-fidelity emulator, rather than as an independent emulator (i.e. as a Latin hypercube sample).
There is some indication that prior information from the LF training samples provides useful information about the best areas of parameter space to sample.
To test this, we split our testing set ($10$ simulations, same resolution as the HF samples) into training and testing sets, then train all $210$ combinations of $6$ samples, and predict the outputs for the remaining $4$ samples.
The error range from this exercise is $2.5-11.5\%$ ($5.5\%$ mean error, $1.5\%$ standard deviation).
Though not a direct comparison, the $3\%$ error we obtain from the HF single-fidelity emulator compares favorably with this, indicating that the HF samples selected are an improvement over using a Latin hypercube sampling scheme.

\begin{figure}
    \centering
	\includegraphics[width=\columnwidth]{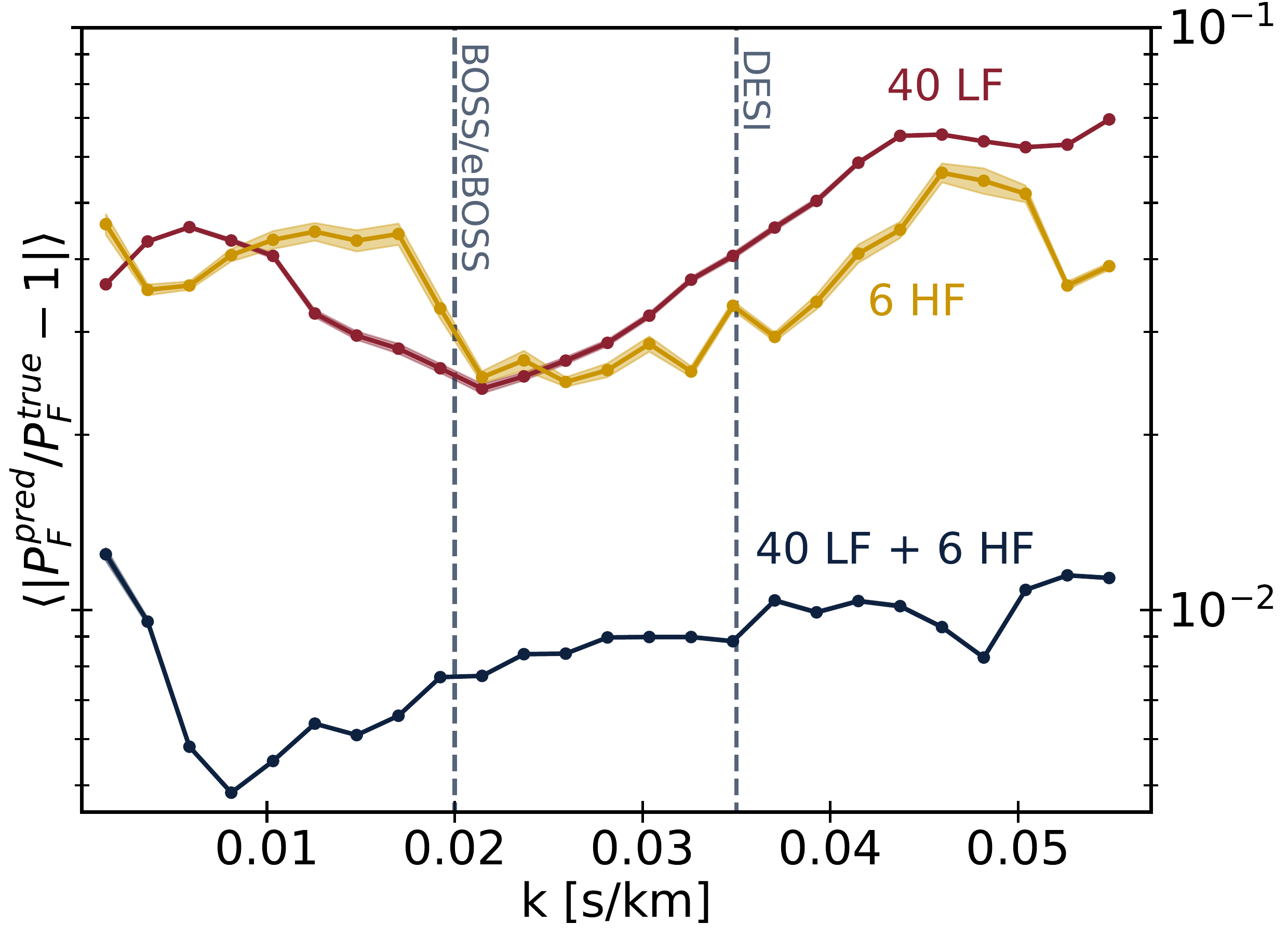}
    \caption{Comparing the prediction error as a function of (linearly binned) wavevector for multi- and single-fidelity emulators.
    This is the mean error across all redshifts and $10$ test simulations.
    The shaded regions are the variance in the prediction error.
    Dashed lines show the highest $k$ probed by BOSS/eBOSS \citep{2019JCAP...07..017C}, and the estimated reach for DESI \citep{2022arXiv220307491V}.}
    \label{fig:k_error}
\end{figure}

Figure~\ref{fig:k_error} shows the mean prediction error, averaged over all redshifts and $10$ test simulation outputs, as a function of wavevector $k$.
In this, and the following figures, the shaded region around the curves is the variance in prediction error ($| P_F^{\text{pred}}/P_F^{\text{true}} - 1 |$), to give a sense of how much the error varies beyond the mean.
The multi-fidelity emulator outperforms the single-fidelity emulators at all scales, with an error between $0.5-1.5\%$.
The LF (HF) single-fidelity emulator has error between $2-7\%$ ($2-6\%$).

Both single-fidelity emulators and the multi-fidelity emulator trend towards higher error for small scales.
The LF emulator dips $1-2\%$ around $k\approx0.02$ s/km.
The dip occurs on scales at which the low resolution flux power spectra goes from overestimating to underestimating the high resolution power (see Figure~\ref{fig:low_high_comp}, for $z\le4.6$).
The uptick in the multi-fidelity emulator error for the largest k-bins is also present in the HF emulator, indicating that there is a scarcity of large scale modes available in the emulator training.

\begin{figure*}
    \centering
	\includegraphics[width=2\columnwidth]{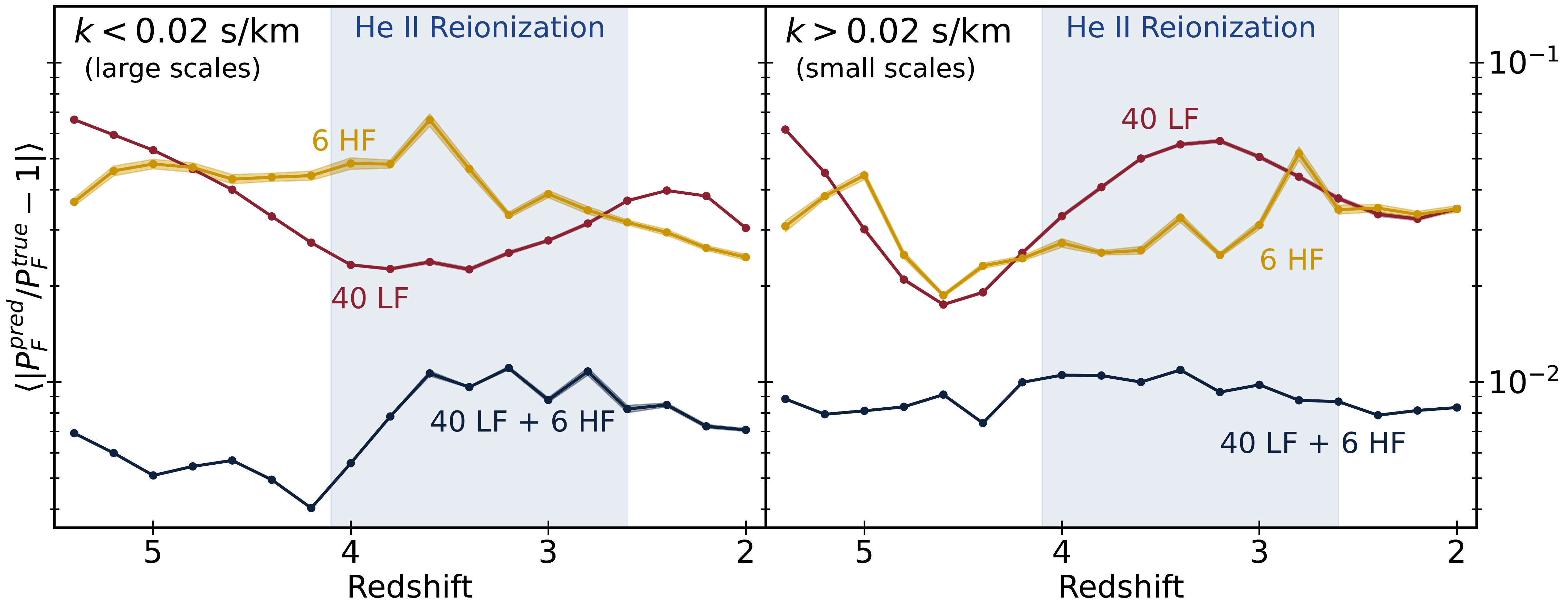}
    \caption{Comparing the prediction error as a function of redshift and scale for multi- and single-fidelity emulators.
    This is mean error across all $10$ test simulations.
    The blue shaded region shows the extent of helium reionization in our simulations (see parameter limits in Figure~\ref{fig:samples}).}
    \label{fig:z_scales_error}
\end{figure*}

Figure~\ref{fig:z_scales_error} shows how the emulators perform as a function of redshift and scale.
In the following, we define small scales as $k > 0.02$ s/km and large scales as $k \le 0.02$ s/km (divided at the smallest scale accessible by BOSS/eBOSS data \citep{2019JCAP...07..017C}).
On large scales (Figure~\ref{fig:z_scales_error}, left panel), the LF emulator error decreases with time until $z\approx4$, then slowly increases.
This trend is because, as can be seen in Figure~\ref{fig:low_high_comp} (left of the BOSS/eBOSS dashed line), the low resolution flux power comes into better agreement with the high resolution flux power as it nears $z\approx4$.
We also found that the LF simulations do not cool as efficiently as the HF simulations after He~{\sc ii} reionization, likely leading to the rise in error for $z\le3.2$.

On small scales (Figure~\ref{fig:z_scales_error}, right panel), the LF emulator error is more variable.
The dip in error seen around $z\approx4.6$ occurs as the low resolution flux power crosses from overestimating to underestimating the high resolution flux power (Figure~\ref{fig:low_high_comp} right of the BOSS/eBOSS dashed line).
The subsequent rise in error is due to the loss of small-scale power and consequent under-estimation of the flux power spectrum in the low-fidelity simulations.

The trends in redshift and scale seen in the LF emulator performance are due not to interpolation error, but to the different numerical resolution of the two simulation fidelities, since this emulator is not predicting the flux power at the higher resolution.
Some differences are connected to temperature differences between the LF and HF simulations.
For low densities ($\sim1$ times the mean density), the LF simulations are colder than the HF simulations at high redshift, but come into better agreement leading up to He~{\sc ii} reionization, after which they diverge from the HF simulations again.
For higher densities ($1-100$ times the mean density), the LF simulations are once again colder than the HF simulations at high redshift, but at lower redshift they are too hot (with a crossover at $z\approx4.6$).
As higher redshifts probe lower densities, the error initially decreases with redshift, before rising again towards the lowest redshifts.

On both large and small scales the HF emulator errors are around $3.5\%$, dominated by sampling variance.
During He~{\sc ii} reionization the HF emulator has more variation in error, which is probably exacerbated by our small box sizes.
The increased variation during He~{\sc ii} reionization further indicates that the primary source of error for the HF emulator is the sample size of the training set.

On small scales, the multi-fidelity emulator error is insensitive to redshift and small ($0.9\%$).
On large scales, the multi-fidelity emulator error slightly decreases until $z=4.2$, then increases with the onset of He~{\sc ii} reionization, before flattening again.
The trend is also more variable during He~{\sc ii} reionization, indicating that emulator finds it more difficult to learn the mapping during this process.
However, the multi-fidelity emulator still outperforms the single-fidelity emulators, with an error between $\approx0.4-1\%$.

\subsection{Emulator Runtime}\label{sec:runtime}

While we have shown that the multi-fidelity emulator outperforms the single-fidelity emulators presented here, it still remains to show that it is more computationally cost efficient.
We could, for example, add more training simulations to our single-fidelity HF emulator and get a similarly accurate high resolution emulator.
However, the computational cost would increase significantly.
By comparing the total emulator runtime to prediction error, we can determine the choice that balances computational cost and accuracy.
In practice, the important question is to determine the computational cost at which a given emulation technique can achieve a desired accuracy.
The computational cost of training the emulators is subdominant ($\mathcal{O}(1)$ cpu-hours) to running the training simulations, so in the following we only consider the runtime for the simulation suites.

\begin{figure}
    \centering
	\includegraphics[width=\columnwidth]{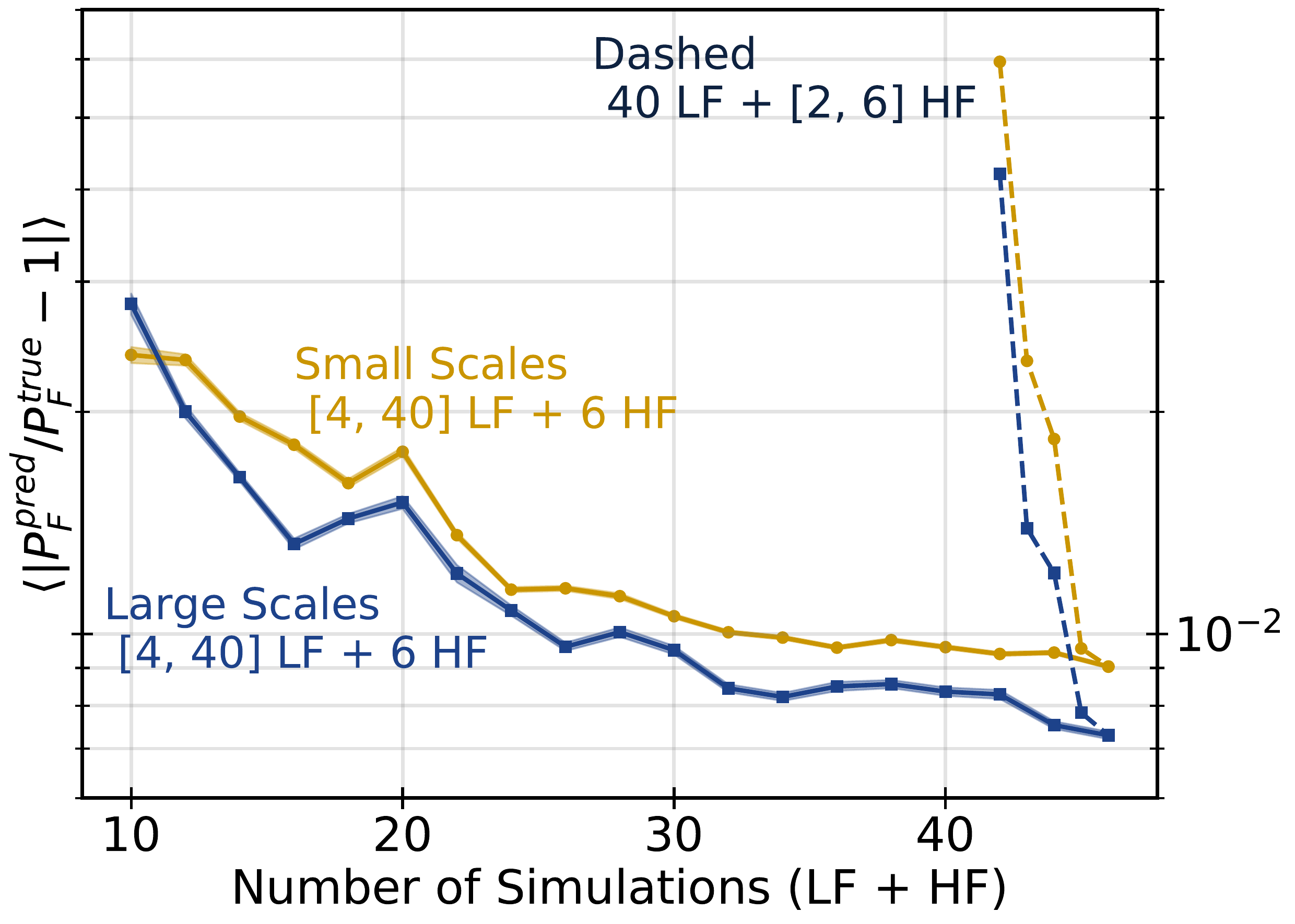}
    \caption{
    Emulator prediction error as a function of the number of simulations used in training the emulator.
    This is the mean error across all redshifts and $10$ test simulation outputs.
    The prediction error is broken down into large ($k < 0.02$ s/km) and small ($k > 0.02$ s/km) scales, as in Figure~\ref{fig:z_scales_error}.
    The solid lines show how the average error depends on the number of LF simulations, while the dashed lines show how the average error depends on the number of HF simulations once all LF simulations are included.
    }
    \label{fig:nsim_scales}
\end{figure}

Figure~\ref{fig:nsim_scales} shows the mean prediction error (averaged over all redshifts and test outputs) as a function of the number of simulations used in the training set, for small and large scales (as defined in Section~\ref{sec:results} and Figure~\ref{fig:z_scales_error}).
The solid lines show prediction errors for multi-fidelity emulators trained using $6$ HF and a varying number of LF simulations.
The small and large scale errors flatten out after $\approx 30$ LF simulations are used in the training set.
The LF simulations allow the emulator to determine how the flux power spectrum depends on the cosmological input parameters, and so this indicates that $30$ LF simulations are needed to explore our $9$ parameter space.
Other emulators range from using $\approx6$ simulations per parameter (e.g. \cite{McClintock:2019, 2022MNRAS.509.2551H}) to $30$ per parameter (e.g. \cite{Euclid:2021}).
The $3-4$ simulations per parameter required here is unusually low, perhaps because the input parameters affect the flux power spectrum close to linearly in much of parameter space.

Dashed lines show prediction errors for multi-fidelity emulators trained using $40$ LF and a varying number of HF simulations.
Adding extra HF simulations to the training set has a larger impact than adding LF simulations.
The addition of each HF simulation generally improves the emulator accuracy for small scales more than for large scales.
This is as expected, since the main purpose of the HF simulations is to learn the mapping from low to high resolution output, with small scales being more resolved in the HF simulations.

\begin{figure}
    \centering
	\includegraphics[width=\columnwidth]{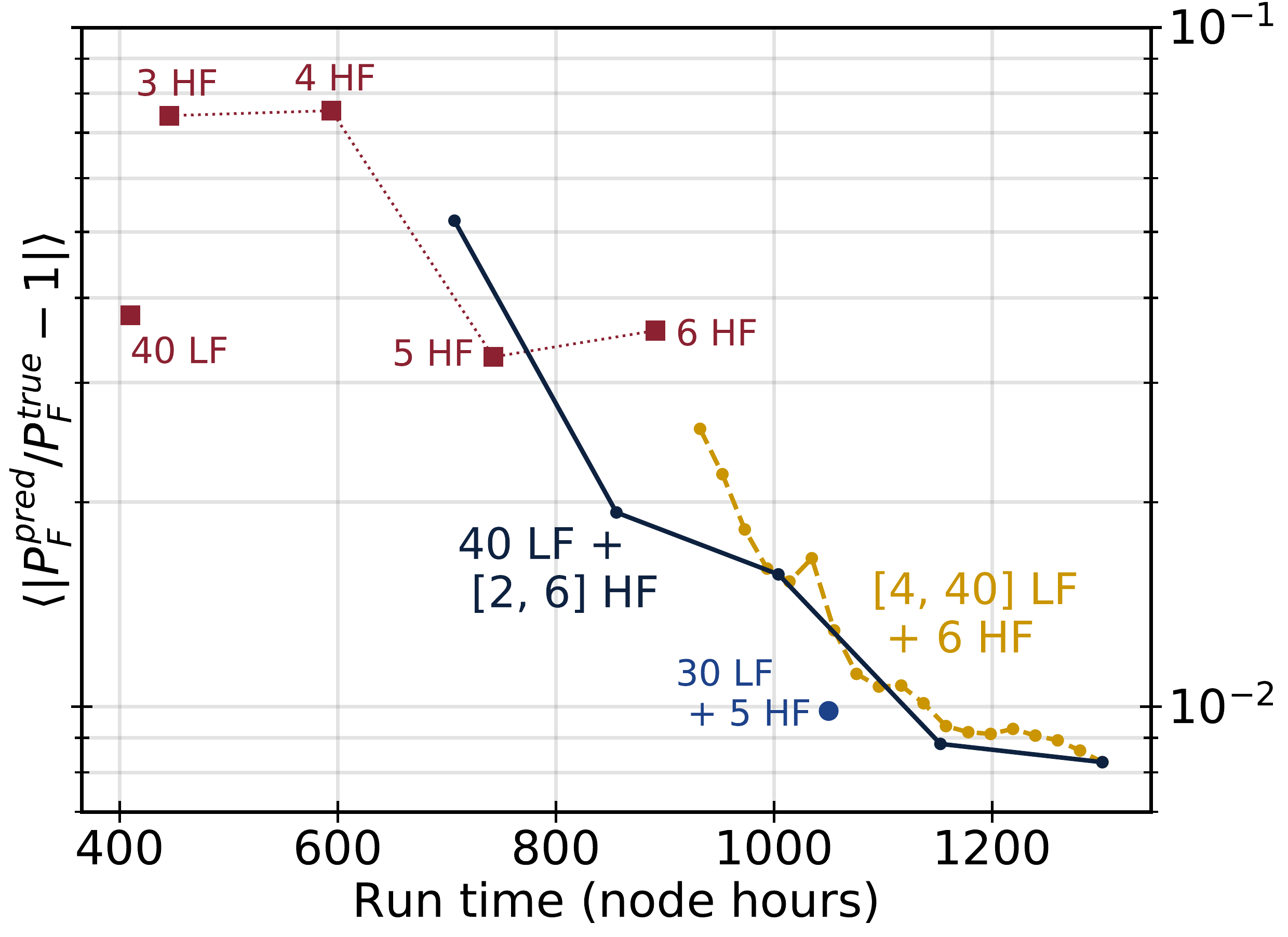}
    \caption{
    Prediction error as a function of total training simulation computational cost.
    This is the mean error for $10$ test simulations over all redshifts and scales.
    The solid line shows the prediction error when changing the number of HF simulations used in training the multi-fidelity emulator.
    Also shown are the single-fidelity emulators (red squares), and the multi-fidelity emulator trend with a varying number of LF training simulations (yellow, dashed).}
    \label{fig:runtime}
\end{figure}

Figure~\ref{fig:runtime} shows the emulator prediction errors as a function of the total runtime (cost of running the training simulations).
All simulations were run on the Frontera supercomputer at the Texas Advanced Computing Center.
The cost is divided between the LF training simulations, which cost $\approx10$ node hours each, and the HF training simulations, which cost $\approx150$ node hours each.

The dashed trend shows the same emulators as Figure~\ref{fig:nsim_scales}, but no longer divided by scale.
Qualitatively it looks the same as both the large scale and small scale results from the previous figure.
The error is flat after $\approx30$ LF training simulations, indicating that a similar accuracy can be achieved using the multi-fidelity emulator with $\approx30$ rather than $40$ LF simulations.
The most efficient $1\%$ error emulator in this study is a multi-fidelity emulator using $30$ LF and $5$ HF training simulations (the cost for this was $\approx1050$ node hours).
The most accurate emulator is the $40$ LF, $6$ HF multi-fidelity emulator, with error $0.8\%$, and cost $\approx1300$ node hours.

The dotted line (squares) shows the error and runtime for the single-fidelity emulators.
Following from the $6$ HF single-fidelity emulator result to the dashed (yellow) line, it can be seen that the addition of just a few LF training simulations quickly improves the accuracy.
It can also be seen that in terms of computational cost, the multi-fidelity emulator is more efficient\footnote{At least for errors less than $4\%$, the approximate amount by which the LF simulations fail to be converged.}.
Note that the HF training simulations are not selected to optimize a single-fidelity emulator, but instead are selected to optimize the multi-fidelity emulator.
They thus use prior information provided by the LF training simulations and so perform better than a naive Latin hypercube construction of a HF emulator using $6$ training samples.
Our multi-fidelity scheme is thus an even larger improvement on a single-fidelity model than Figure~\ref{fig:runtime} suggests.

The solid line shows the error and runtime for the multi-fidelity emulator trained using $40$ LF simulations, and $2-6$ HF simulations.
The point on the solid line corresponding to $40$ LF, $2$ HF has a similar cost, but slightly worse performance than the $5$ HF single-fidelity result.
Adding a third HF training sample decreases the error more for the multi-fidelity emulator (error for $40$ LF, $3$ HF emulator) than it does for the single-fidelity emulator (error marked $6$ HF).
Adding a $6$th HF simulation to the $40$ LF, $5$ HF multi-fidelity emulator produces a relatively small improvement in error, perhaps indicating that stochasticity in the simulations due to our relatively small box size, is beginning to dominate over interpolation error.
\section{Conclusions}\label{sec:conclusions}

In this work we developed and tested a multi-fidelity emulator for the simulated Lyman-$\alpha$ forest flux power spectrum.
Emulators address the growing computational demands of simulations, which must be run at increasingly high resolutions to allow analysis of the increasing quality and quantity of observational data.
Here, we use a Gaussian process based emulator that addresses this demand by, in a Bayesian framework, training an interpolating function to predict the output (Lyman-$\alpha$ forest flux power spectrum) for a given input (simulation input parameters).
Relatively few simulations are required to accurately predict across the span of input parameter space, making emulators especially useful for parameter inference problems.

The multi-fidelity framework allows a further reduction in computational cost by dividing the emulator training samples into multiple (in our case two) fidelities.
The low-fidelity (low-resolution) training samples allow the emulator to learn how the the outputs depend on input parameters.
The high-fidelity (high-resolution) training samples correct numerical errors in the low-fidelity emulator with a (parameter-dependent) mapping from low- to high-fidelity.
Thus, the emulator can be trained with a large sample of low-fidelity training simulations and a small subset of high-fidelity training simulations.

Our training suite included $40$ low resolution hydrodynamical simulations ($30$ Mpc/h simulation box length, $256^3$ particles) and $6$ high resolution hydrodynamical simulations ($30$ Mpc/h simulation box length, $512^3$ particles).
Using the Lyman-$\alpha$ forest flux power spectrum extracted from these simulations, we trained single- and multi-fidelity emulators to predict the high resolution flux power spectrum.
Ten independent simulations were run to test the prediction accuracy of the trained emulators.

In summary, the multi-fidelity emulator:
\begin{itemize}
    \item Modelled a redshift range $5.4 - 2$ (in $18$ redshift bins with $\Delta z=0.2$), on scales ranging from $k=1.4\times10^{-3}$ to $k=5.7\times10^{-2}$ s/km (in $25$ bins).
    \item Achieved sub-$1\%$ error on most scales and redshifts when averaged over $10$ test simulations.
    \item Reached an average error of $0.8\%$.
    \item Achieved $1\%$ average error most cost efficiently using a training set with $30$ low resolution and $5$ high resolution simulations.
\end{itemize}

The low resolution single-fidelity emulator ($4\%$ average error) predicts the low resolution flux power, so it is limited by real differences between the output of the two resolutions.
The high resolution single-fidelity emulator ($3\%$ average error) is limited by the small number of training samples.
It is likely that the average error for the high resolution single-fidelity emulator could be improved to match the multi-fidelity emulator performance with the addition of more training simulations.
However, the high resolution single-fidelity emulator quickly increases in computational cost with additional samples, and we expect it would thus be more expensive than our multi-fidelity emulator.

Some important caveats to our results are that the Lyman-$\alpha$ forest is converged at the $\approx5\%$ level in our high resolution simulations, and the box size is small.
In a forthcoming work, a model that uses two different box sizes (rather than two different resolutions) to construct a multi-fidelity emulator will be developed and tested.
While there is no direct evidence to suggest that changing the resolution or box size would significantly enhance or diminish the accuracy of the emulators presented here, it still remains to be tested on simulations with higher resolution and larger box sizes.
In a forthcoming work, we test the multi-fidelity framework on larger box size, higher resolution simulations, and use this multi-fidelity emulator for cosmological inference.

\section*{Acknowledgements}
MAF is supported by a National Science Foundation Graduate Research Fellowship under grant No. DGE-1326120.
MFH is supported by a National Aeronautics and Space Administration FINESST under grant No. ASTRO20-0022.
SB is supported by NSF grant AST-1817256.

Computing resources were provided by Frontera LRAC AST21005.
The authors acknowledge the Frontera computing project at the Texas Advanced Computing Center (TACC) for providing HPC and storage resources that have contributed to the research results reported within this paper.
Frontera is made possible by National Science Foundation award OAC-1818253.
URL: \url{http://www.tacc.utexas.edu}

\section*{Data Availability}
Flux power spectra generated from the low resolution, high resolution, and testing sets are available at \url{https://github.com/mafern/MFEmulatorLyaData}.
HDF5 and plain text (appropriate for multi-fidelity emulation) formats are available.
Select single- and multi-fidelity emulator predictions for the $10$ testing simulations are also available from the same repository.
The spectra underlying the flux power are available upon request.

\bibliographystyle{mnras}
\bibliography{refs}

\appendix
\section{Non-linear Multi-Fidelity Emulator}\label{sec:nonlinear}

\begin{figure*}
    \centering
	\includegraphics[width=2\columnwidth]{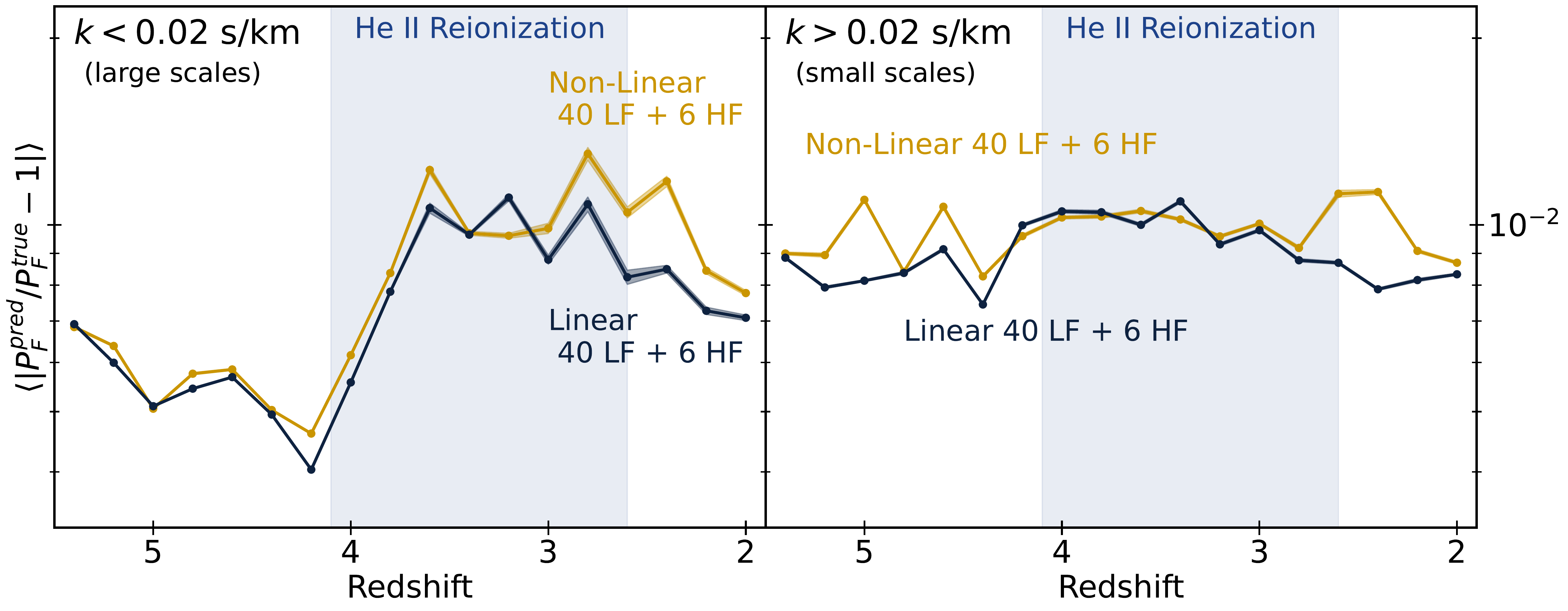}
    \caption{Comparing the prediction error as a function of redshift and scale for linear and non-linear multi-fidelity emulators.
    This is mean error across all $10$ test simulations.
    The shaded region shows the extent of helium reionization in our simulations (see parameter limits in Figure~\ref{fig:samples}).}
    \label{fig:nonlinear}
\end{figure*}

In the main text we have explored the effectiveness of a linear multi-fidelity emulator (the KO model, or AR1).
In the linear model, the mapping from LF to HF is $f_{\text{HF}}(\thetavec) = \rho f_{\text{LF}}(\thetavec) + \delta(\thetavec),$ where $f_{\text{HF}}$ and $f_{\text{LF}}$ are the emulator predictions at those resolutions, and $\rho$ is independent of the input parameters $\thetavec$.

Here, we compare those results with the results using a non-linear multi-fidelity emulator (non-linear autoregressive GP, or NARGP).
In the non-linear multi-fidelity model, proposed by \cite{2017RSPSA.47360751P}, the mapping is a function of both the LF output and the input parameters.
We model this as:
$$f_{\text{HF}}(\thetavec) = \rho(\thetavec, \tilde{f}_{\text{LF}}(\thetavec)) + \delta(\thetavec),$$ 
such that $\rho$ depends on both the input parameters and LF posterior output.
The LF outputs, as is the case with the linear model, are median normalized such that the assumption on the Gaussian process of zero mean is more reasonable, $\tilde{f}_{\text{LF}}(\thetavec) = f_{\text{LF}}(\thetavec)/\mu_{\text{LF}} - 1$.

Following \cite{2017RSPSA.47360751P}, $\rho$ is modelled as a Gaussian process with input from both input cosmologies for HF, $\thetavec_\mathrm{HF}$, and the output from LF, $\tilde{f}_{\text{LF}}(\thetavec)$.
The NARGP construction results in a deep Gaussian process model \citep{pmlr-v31-damianou13a}.
We follow the approximation in \cite{2017RSPSA.47360751P} and replace $\tilde{f}_{\text{LF}}(\thetavec)$ with its posterior distribution.
Thus, the training reduces to training two regular GPs recursively.

In Figure~\ref{fig:nonlinear} we show the prediction errors separated into small and large scales, as a function of redshift for both the linear and non-linear multi-fidelity emulators.
They perform similarly, with the linear emulator being more accurate at low redshifts on all scales, and high redshifts for small scales.
The difference between the average error for the linear and non-linear models (over all scales and redshifts) is $0.08\%$.
This is in contrast to emulation of the matter power spectrum in \cite{2022MNRAS.509.2551H}, where the non-linear model outperformed the linear model.

It is worth noting that the non-linear model agrees closely with the linear model when using the full suite of training simulations, but lags behind the linear model when using fewer HF training simulations.
For example, the difference in the average error between linear and non-linear models using $40$ LF and $3$ HF is $\approx2\%$ ($1.9\%$ error for linear, $3.7\%$ error for non-linear).
When using $4$ HF the difference is $\approx1\%$ ($1.5\%$, $2.5\%$), and when using $5$ HF the difference is $\approx0.4\%$ ($0.9\%$, $1.3\%$).
While differences in the effectiveness of the non-linear model may be due to the quantity being emulated (matter power versus flux power), one likely reason for the difference is the number of input parameters.
In \cite{2022MNRAS.509.2551H}, five input parameters are used, while in this work we use nine.
The non-linear model uses the posterior of the LF output, which requires Monte-Carlo sampling.
It is possible that the additional dimensions degrade the performance of the Monte-Carlo integration, and thus the performance of the non-linear model.
One other reason may be the larger number of hyperparameters that need to be optimized in the training.

\bsp
\label{lastpage}
\end{document}